\documentclass[notitlepage,a4paper,aps,prd,onecolumn,superscriptaddress,nofootinbib,groupedaddress]{revtex4}

\usepackage{amsmath,comment}
\usepackage{amsfonts,color}
\usepackage{amsmath}
\usepackage{amsthm}
\usepackage{graphicx} 
\usepackage{pdfpages}
\usepackage{amsmath}
\usepackage{empheq}
\usepackage{amsfonts,color}
\usepackage{amssymb,float}
\usepackage{physics}
\usepackage{dsfont}
\usepackage{booktabs,multirow}
\usepackage{titlesec}

\titleformat{\paragraph}
{\normalfont\normalsize\bfseries}{\theparagraph}{1em}{}
\titlespacing*{\paragraph}
{0pt}{3.25ex plus 1ex minus .2ex}{1.5ex plus .2ex}

\usepackage[none]{hyphenat}

\usepackage{amsmath}
\usepackage[utf8]{inputenc}
\allowdisplaybreaks
\usepackage{color}
\usepackage{hyperref}
\hypersetup{
    colorlinks=true,
    linkcolor=blue,
    filecolor=magenta,      
   citecolor=blue
}

\usepackage{enumitem}

\usepackage{tikz}
\usetikzlibrary{shapes.geometric}
\usetikzlibrary{arrows.meta,arrows}

\begin{document}
\title{Black hole superradiance in Poincaré gauge theory}

\author{Sebastian Bahamonde}
\email{sbahamondebeltran@gmail.com}
\affiliation{Cosmology, Gravity, and Astroparticle Physics Group, Center for Theoretical Physics of the Universe,
Institute for Basic Science (IBS), Daejeon, 34126, Korea.}

\author{Jorge Gigante Valcarcel}
\email{jorgevalcarcel@ibs.re.kr}
\affiliation{Center for Geometry and Physics, Institute for Basic Science (IBS), Pohang 37673, Korea.}

\begin{abstract}

We investigate the phenomenon of black hole superradiance in the presence of torsion within the framework of Poincaré gauge theory. In particular, in contrast to the classical approach of General Relativity, we show that the inclusion of torsion in the space-time geometry enables the energy extraction from rotating black holes by Dirac fermions via chiral asymmetry, while preserving the Pauli exclusion principle.

\end{abstract}

\maketitle

\section{Introduction}

Black holes are among the most remarkable predictions of General Relativity (GR), arising naturally as solutions of the Einstein's field equations with a significant degree of symmetry~\cite{Stephani:2003tm,Griffiths:2009dfa}. Notably, under stationary and axisymmetric conditions, the Kerr solution describes the space-time geometry of a rotating black hole~\cite{Kerr:1963ud}, while further generalisations incorporate the gravitational effects provided by the energy-momentum tensor of the electromagnetic and Yang-Mills fields~\cite{Newman:1965tw,Newman:1965my,Volkov:1997qb,Volkov:1998cc,Kleihaus:2000kg}.

Although historically considered perfectly absorbing systems, black holes can in fact transfer energy to surrounding fields through a variety of classical and quantum processes~\cite{Penrose:1969pc,Christodoulou:1970wf,Hawking:1974rv,Hawking:1975vcx}. Indeed, the presence of an event horizon together with an ergoregion allows rotating black holes to behave as dissipative systems that lead to superradiant amplification of bosonic fields~\cite{Zeldovich:1971ffh,Zeldovich:1972zqp,Press:1972zz,Starobinskii:1973vzb,Teukolsky:1974yv,East:2013mfa,Brito:2015oca,East:2017ovw}. However, early analyses of the Dirac equation in rotating black holes showed that classical Dirac fields do not exhibit superradiance~\cite{Unruh:1973bda,Chandrasekhar:1976ap,Maeda:1976tm,Wagh:1985lyl,Dolan:2015eua}. Thereby, unlike bosonic fields, the scattered fermionic radiation is never amplified and no energy can be extracted from the black hole, while spontaneous particle production can occur in the quantum regime~\cite{Unruh:1974bw,Dai:2023zcj}.

Following these lines, in this work we analyse the role of a space-time torsion in the behaviour of Dirac fermions in rotating black holes. In particular, we show that the interaction mediated by torsion enables the energy extraction from black holes without inducing wave amplification, thus maintaining consistency with the Pauli exclusion principle. For this purpose, we consider a slowly rotating black hole recently found in the framework of Poincaré gauge theory, which includes a spin charge as a source of torsion and displays a spin-orbit interaction in the gravitational action~\cite{Bahamonde:2025fei}. Indeed, the presence of a nontrivial axial mode for torsion in the solution plays a central role in our study. First, although the torsion tensor can be decomposed into three irreducible parts under the four-dimensional pseudo-orthogonal group, only its axial mode mediates the interaction with Dirac fermions, as prescribed by the minimal coupling principle~\cite{Hehl:1971qi}. Furthermore, the parity-odd nature of this mode induces opposite frequency shifts in the respective helicity states of the Dirac fermion, which turns out to be crucial in the process of energy extraction.

This paper is organised as follows. First, in Sec.~\ref{sec:model} we briefly introduce the rotating black hole solution which will be considered in our calculations. Then, in Sec.~\ref{sec:DiracEq} we analyse the Dirac equation in the presence of torsion and determine the condition that the axial mode of the solution must satisfy to guarantee separability. In that case, the underlying system of equations can be readily integrated, which allows us to obtain analytical solutions near the event horizon, as discussed in Sec.~\ref{sec:sol}. Focusing on the ingoing modes of the Dirac fermion, in Sec.~\ref{sec:results} we compute then the conserved current at the horizon and show that the associated net number current is strictly positive, reflecting the absence of wave amplification. In addition, we also evaluate the weak-energy condition for a timelike projection of the energy-momentum tensor of the Dirac field, which turns out to be violated for a specific range of frequencies that includes corrections of the axial mode of torsion. A further calculation of the energy current reveals that the shift of frequencies provided by torsion allows this quantity to take negative values, thus signaling the energy extraction by the Dirac fermion from the rotating black hole. Finally, we present our conclusions in Sec.~\ref{sec:conclusions}.

We work in natural units $c=G=\hbar=1$ and consider the metric signature $(+,-,-,-)$. In addition, we use a tilde accent to denote quantities defined from the affine connection with torsion. On the other hand, Latin and Greek indices run from $0$ to $3$, referring to anholonomic and coordinate bases, respectively.

\section{Rotating black holes and gravitational spin-orbit interaction in Poincaré gauge theory}\label{sec:model}

The extension of GR towards a post-Riemannian geometry with torsion allows the study of the physical implications of the intrinsic spin of matter in the space-time~\cite{Kibble:1961ba,sciama1962analogy,Sciama:1964wt}. The gravitational field characterised by curvature and torsion is then described by a gauge field associated with the external rotations and translations of the Poincaré group~\cite{Hehl:1976kj,Obukhov:1987tz,Blagojevic:2013xpa,ponomarev2017gauge,Obukhov:2022khx}. Thereby, a gauge connection valued in the Lie algebra of the Poincar\'{e} group can be introduced to describe the gravitational interaction:
\begin{equation}
    A_{\mu}=e^{a}{}_{\mu}P_{a}+\omega^{a b}{}_{\mu}J_{a b}\,,
\end{equation}
where $e^{a}{}_{\mu}$ and $\omega^{a b}{}_{\mu}$ respectively represent the tetrad field and the spin connection, satisfying the following relations with the metric tensor $g_{\mu\nu}$ and the affine connection $\tilde{\Gamma}^{\lambda}{}_{\rho\mu}$ of a Riemann-Cartan space-time:
\begin{align}
    g_{\mu \nu}&=e^{a}{}_{\mu}\,e^{b}{}_{\nu}\,\eta_{a b}\,,\\
    \label{anholonomic_connection}
    \omega^{a b}{}_{\mu}&=e^{a}{}_{\lambda}\,e^{b\rho}\,\tilde{\Gamma}^{\lambda}{}_{\rho \mu}+e^{a}{}_{\lambda}\,\partial_{\mu}\,e^{b\lambda}\,,
\end{align}
with $\eta_{a b}$ the local Minkowski metric. On the other hand, $P_{a}$ and $J_{a b}$ correspond to the generators of the Poincar\'{e} group, satisfying commutation relations:
\begin{align}
    \left[P_{a},P_{b}\right]&=0\,,\\
    \left[P_{a},J_{bc}\right]&=i\,\eta_{a[b}\,P_{c]}\,,\\
    \left[J_{ab},J_{cd}\right]&=\frac{i}{2}\,\left(\eta_{ad}\,J_{bc}+\eta_{cb}\,J_{ad}-\eta_{db}\,J_{ac}-\eta_{ac}\,J_{bd}\right).
\end{align}

Thus, the corresponding field strength tensor derived from the gauge connection includes translational and rotational parts: 
\begin{align}
    F^{a}{}_{\mu\nu}&=\partial_{\mu}e^{a}{}_{\nu}-\partial_{\nu}e^{a}{}_{\mu}+\omega^{a}{}_{b\mu}\,e^{b}{}_{\nu}-\omega^{a}{}_{b\nu}\,e^{b}{}_{\mu}\,,\\
    F^{ab}{}_{\mu\nu}&=\partial_{\mu}\omega^{a b}{}_{\nu}-\partial_{\nu}\omega^{a b}{}_{\mu}+\omega^{a}{}_{c\mu}\,\omega^{c b}{}_{\nu}-\omega^{a}{}_{c\nu}\,\omega^{c b}{}_{\mu}\,,
\end{align}
which are indeed related to the torsion and curvature tensors as follows:
\begin{align}
    F^{a}\,_{\mu\nu}&=e^{a}\,_{\lambda}T^{\lambda}\,_{\nu\mu}\,,
\\
    F^{a b}{}_{\mu\nu}&=e^{a}\,_{\lambda}e^{b\rho}\tilde{R}^{\lambda}\,_{\rho\mu\nu}\,,
\end{align}
where
\begin{align}
    T^{\lambda}\,_{\mu \nu}&=2\tilde{\Gamma}^{\lambda}\,_{[\mu \nu]}\,,\\
    \tilde{R}^{\lambda}\,_{\rho \mu \nu}&=\partial_{\mu}\tilde{\Gamma}^{\lambda}\,_{\rho \nu}-\partial_{\nu}\tilde{\Gamma}^{\lambda}\,_{\rho \mu}+\tilde{\Gamma}^{\lambda}\,_{\sigma \mu}\tilde{\Gamma}^{\sigma}\,_{\rho \nu}-\tilde{\Gamma}^{\lambda}\,_{\sigma \nu}\tilde{\Gamma}^{\sigma}\,_{\rho \mu}\,.
\end{align}

In this context, the study of rotating black holes with torsion is especially relevant, since the interaction provided by their intrinsic and extrinsic angular momentum parameters can have significant effects at macroscopical scales. Indeed, a rotating black hole solution displaying this interaction in the gravitational action was recently obtained for a particular model of cubic Poincaré gauge theory~\cite{Bahamonde:2025fei}. Specifically, due to the computational intractability of the field equations of the theory, the solution arising from this model reduces to the slowly rotating Kerr black hole:
\begin{align}\label{eq:metric}
    ds^2=&\;\Psi(r)dt^2-\frac{1}{\Psi(r)}d r^2-r^2 d\vartheta^2- r^2 \sin^2\vartheta \,d\varphi^2+2a\bigl(1-\Psi(r)\bigr)\sin ^2\vartheta\, dt  d\varphi\,,
\end{align}
while the spin charge and the spin-orbit interaction are encoded in the axial mode of torsion:
\begin{align}
    S_{t}(r,\vartheta)&=-\,\Psi(r)S_{r}(r,\vartheta) = -\,\frac{6N_{1}\kappa_{\rm s}}{r}+\frac{2a}{r^{2}\Psi(r)}\left[6r\Psi(r)F(r,\vartheta)+\left(4\Psi(r)-3\right)\cos\vartheta\right],\label{AxialComp1}\\
    S_{\vartheta}(r,\vartheta) &=0\,, \quad S_{\varphi}(r,\vartheta) =\frac{6aN_{1}\kappa_{\rm s}}{r}\sin^{2}\vartheta\,,\label{AxialComp2}
\end{align}
with
\begin{equation}
    \Psi(r)=1-\frac{2m}{r}\,.
\end{equation}

In fact, from a physical point of view, the presence of this mode in the solution is especially relevant, since the interaction between torsion and Dirac fermions under minimal coupling occurs exclusively through this mode~\cite{Hehl:1971qi}.

\section{Dirac equation in the presence of torsion}\label{sec:DiracEq}

Considering the irreducible decomposition of the torsion tensor under the four-dimensional pseudo-orthogonal group~\cite{McCrea:1992wa}:
\begin{equation}
    T^{\lambda}{}_{\mu \nu}=\frac{1}{3}\left(\delta^{\lambda}{}_{\nu}T_{\mu}-\delta^{\lambda}{}_{\mu}T_{\nu}\right)+\frac{1}{6}\,\varepsilon^{\lambda}{}_{\rho\mu\nu}S^{\rho}+t^{\lambda}{}_{\mu \nu}\,,
\end{equation}
with
\begin{align}\label{Tdec1}
    T_{\mu}&=T^{\nu}{}_{\mu\nu}\,,\\
    S_{\mu}&=\varepsilon_{\mu\lambda\rho\nu}T^{\lambda\rho\nu}\,,\\
    t_{\lambda\mu\nu}&=T_{\lambda\mu\nu}-\frac{2}{3}g_{\lambda[\nu}T_{\mu]}-\frac{1}{6}\,\varepsilon_{\lambda\rho\mu\nu}S^{\rho}\,,\label{Tdec3}
\end{align}
the Dirac equation describing a fermion minimally coupled to torsion reads~\cite{Hehl:1971qi}:
\begin{equation}
    \gamma^{\mu}\partial_{\mu}\psi +\gamma^{\mu}\omega_{\mu}\psi-\frac{i}{4}\gamma^{5}\gamma^{\mu}S_{\mu}\psi+i\mu\psi=0\,.
\end{equation}
Hence, as previously mentioned, the interaction between the Dirac field and torsion under minimal coupling is mediated solely by the axial mode.

For simplicity in the calculations, henceforth we focus on the massless case $\mu = 0\,$. Then, by taking into account the Weyl representation
\begin{eqnarray}
\gamma^{0}=\left(
\begin{array}{cc}
\mathbf{0} & \mathds{1} \\
\mathds{1} & \mathbf{0}
\end{array} \right),\quad \gamma^{i}=\left(
\begin{array}{cc}
\mathbf{0} & \sigma^{i} \\
-\,\sigma^{i} & \mathbf{0}
\end{array} \right),
\end{eqnarray}
where $\{\sigma^{i}\}_{i=1}^{3}$ are the three Pauli matrices
\begin{eqnarray}
\sigma^{1}=\left(
\begin{array}{cc}
0 & 1 \\
1 & 0
\end{array} \right),\quad \sigma^{2}=\left(
\begin{array}{cc}
0 & -i \\
i & 0
\end{array} \right),\quad \sigma^{3}=\left(
\begin{array}{cc}
1 & 0 \\
0 & -1
\end{array} \right),
\end{eqnarray}
and expressing the spinor field as
\begin{eqnarray}
\psi(t,r,\vartheta,\varphi)=e^{i\left(k\varphi-\omega t\right)}\,r^{-\,1/2}\,\Psi^{-\,1/4}(r)\left(
\begin{array}{c}
\varrho^{\,-1/2}(r,\vartheta)\,\eta_{-}(r,\vartheta) \\
\bar{\varrho}^{\,-1/2}(r,\vartheta)\,\eta_{+}(r,\vartheta)
\end{array} \right), \quad \eta_{\pm}(r,\vartheta)=\left(
\begin{array}{c}
\eta_{\pm}^{(1)}(r,\vartheta) \\
\eta_{\pm}^{(2)}(r,\vartheta)
\end{array} \right),
\end{eqnarray}
with
\begin{equation}
    \varrho(r,\vartheta) = r+ia\cos\vartheta\,, \quad \bar{\varrho}(r,\vartheta) = r-ia\cos\vartheta\,,
\end{equation}
the four components of the Dirac equation acquire the following form in the slowly rotating Kerr black hole endowed with the axial mode of torsion:
\begin{align}
    &\frac{i}{r\sqrt{\Psi(r)}}\left(ak-\omega r^2\right)\eta_{+}^{(1)}(r,\vartheta)+r\sqrt{\Psi(r)}\,\partial_{r}\eta_{+}^{(2)}(r,\vartheta)-\frac{i}{2}\left(2\partial_{\vartheta}+\cot{\vartheta}\right)\eta_{+}^{(2)}(r,\vartheta)+i\left(k\csc{\vartheta}-a\omega\sin\vartheta\right)\eta_{+}^{(1)}(r,\vartheta)\nonumber\\
    &+\frac{i}{4r\sqrt{\Psi(r)}}\left(r^{2}S_{t}(r,\vartheta)+aS_{\varphi}(r,\vartheta)\right)\eta_{+}^{(1)}(r,\vartheta)+\frac{ir\sqrt{\Psi(r)}}{4}S_{r}(r,\vartheta)\eta_{+}^{(2)}(r,\vartheta)+\frac{1}{4}S_{\vartheta}(r,\vartheta)\eta_{+}^{(2)}(r,\vartheta)\nonumber\\
    &+\frac{i}{4}\left(a\sin\vartheta S_{t}(r,\vartheta)+\csc{\vartheta}S_{\varphi}(r,\vartheta)\right)\eta_{+}^{(1)}(r,\vartheta)=0\,,\label{DiracTorsComp1}\\
    &\frac{i}{r\sqrt{\Psi(r)}}\left(ak-\omega r^{2}\right)\eta_{+}^{(2)}(r,\vartheta)+r\sqrt{\Psi(r)}\,\partial_{r}\eta_{+}^{(1)}(r,\vartheta)+\frac{i}{2}\left(2\partial_{\vartheta}+\cot{\vartheta}\right)\eta_{+}^{(1)}(r,\vartheta)-i\left(k\csc{\vartheta}-a\omega\sin\vartheta\right)\eta_{+}^{(2)}(r,\vartheta)\nonumber\\
    &+\frac{i}{4r\sqrt{\Psi(r)}}\left(r^{2}S_{t}(r,\vartheta)+aS_{\varphi}(r,\vartheta)\right)\eta_{+}^{(2)}(r,\vartheta)+\frac{ir\sqrt{\Psi(r)}}{4}S_{r}(r,\vartheta)\eta_{+}^{(1)}(r,\vartheta)-\frac{1}{4}S_{\vartheta}(r,\vartheta)\eta_{+}^{(1)}(r,\vartheta)\nonumber\\
    &-\frac{i}{4}\left(a\sin\vartheta S_{t}(r,\vartheta)+\csc{\vartheta}S_{\varphi}(r,\vartheta)\right)\eta_{+}^{(2)}(r,\vartheta)=0\,,\label{DiracTorsComp2}\\
    &\frac{i}{r\sqrt{\Psi(r)}}\left(ak-\omega r^{2}\right)\eta_{-}^{(1)}(r,\vartheta)-r\sqrt{\Psi(r)}\,\partial_{r}\eta_{-}^{(2)}(r,\vartheta)+\frac{i}{2}\left(2\partial_{\vartheta}+\cot{\vartheta}\right)\eta_{-}^{(2)}(r,\vartheta)-i\left(k\csc{\vartheta}-a\omega\sin\vartheta\right)\eta_{-}^{(1)}(r,\vartheta)\nonumber\\
    &-\frac{i}{4r\sqrt{\Psi(r)}}\left(r^{2}S_{t}(r,\vartheta)+aS_{\varphi}(r,\vartheta)\right)\eta_{-}^{(1)}(r,\vartheta)+\frac{ir\sqrt{\Psi(r)}}{4}S_{r}(r,\vartheta)\eta_{-}^{(2)}(r,\vartheta)+\frac{1}{4}S_{\vartheta}(r,\vartheta)\eta_{-}^{(2)}(r,\vartheta)\nonumber\\
    &+\frac{i}{4}\left(a\sin\vartheta S_{t}(r,\vartheta)+\csc{\vartheta}S_{\varphi}(r,\vartheta)\right)\eta_{-}^{(1)}(r,\vartheta)=0\,,\label{DiracTorsComp3}\\
    &\frac{i}{r\sqrt{\Psi(r)}}\left(ak-\omega r^{2}\right)\eta_{-}^{(2)}(r,\vartheta)-r\sqrt{\Psi(r)}\,\partial_{r}\eta_{-}^{(1)}(r,\vartheta)-\frac{i}{2}\left(2\partial_{\vartheta}+\cot{\vartheta}\right)\eta_{-}^{(1)}(r,\vartheta)+i\left(k\csc{\vartheta}-a\omega\sin\vartheta\right)\eta_{-}^{(2)}(r,\vartheta)\nonumber\\
    &-\frac{i}{4r\sqrt{\Psi(r)}}\left(r^{2}S_{t}(r,\vartheta)+aS_{\varphi}(r,\vartheta)\right)\eta_{-}^{(2)}(r,\vartheta)+\frac{ir\sqrt{\Psi(r)}}{4}S_{r}(r,\vartheta)\eta_{-}^{(1)}(r,\vartheta)-\frac{1}{4}S_{\vartheta}(r,\vartheta)\eta_{-}^{(1)}(r,\vartheta)\nonumber\\
    &-\frac{i}{4}\left(a\sin\vartheta S_{t}(r,\vartheta)+\csc{\vartheta}S_{\varphi}(r,\vartheta)\right)\eta_{-}^{(2)}(r,\vartheta)=0\,.\label{DiracTorsComp4}
\end{align}

Thus, changing variables to
\begin{equation}
    H^{(1)}_{\pm}(r,\vartheta)=\eta_{\pm}^{(1)}(r,\vartheta)+\eta_{\pm}^{(2)}(r,\vartheta)\,, \quad H^{(2)}_{\pm}(r,\vartheta)=i\bigl(\eta_{\pm}^{(1)}(r,\vartheta)-\eta_{\pm}^{(2)}(r,\vartheta)\bigr)\,,
\end{equation}
the system of equations~\eqref{DiracTorsComp1}-\eqref{DiracTorsComp4} reads:
\begin{align}
    &r\sqrt{\Psi(r)}\,\partial_{r}H_{+}^{(1)}(r,\vartheta)+\frac{i}{4r\sqrt{\Psi(r)}}\left[4\left(ak-\omega r^2\right)+r^2\left(S_{t}(r,\vartheta)+\Psi(r)S_{r}(r,\vartheta)\right)+aS_{\varphi}(r,\vartheta)\right]H_{+}^{(1)}(r,\vartheta)\nonumber\\
    &+\partial_{\vartheta}H_{+}^{(2)}(r,\vartheta)+\frac{1}{4}\left[2\cot{\vartheta}+4\left(k\csc{\vartheta}-a\omega\sin\vartheta\right)+iS_{\vartheta}(r,\vartheta)+\left(aS_{t}(r,\vartheta)\sin\vartheta+S_{\varphi}(r,\vartheta)\csc{\vartheta}\right)\right]H_{+}^{(2)}(r,\vartheta)=0\,,\label{DiracTorsCompNewVar1}\\
    &r\sqrt{\Psi(r)}\,\partial_{r}H_{+}^{(2)}(r,\vartheta)-\frac{i}{4r\sqrt{\Psi(r)}}\left[4\left(ak-\omega r^2\right)+r^{2}\left(S_{t}(r,\vartheta)-\Psi(r)S_{r}(r,\vartheta)\right)+aS_{\varphi}(r,\vartheta)\right]H_{+}^{(2)}(r,\vartheta)\nonumber\\
    &-\partial_{\vartheta}H_{+}^{(1)}(r,\vartheta)-\frac{1}{4}\left[2\cot{\vartheta}-4\left(k\csc{\vartheta}-a\omega\sin\vartheta\right)+iS_{\vartheta}(r,\vartheta)-\left(aS_{t}(r,\vartheta)\sin\vartheta+S_{\varphi}(r,\vartheta)\csc{\vartheta}\right)\right]H_{+}^{(1)}(r,\vartheta)=0\,,\label{DiracTorsCompNewVar2}\\
    &r\sqrt{\Psi(r)}\,\partial_{r}H_{-}^{(1)}(r,\vartheta)-\frac{i}{4r\sqrt{\Psi(r)}}\left[4\left(ak-\omega r^2\right)-r^{2}\left(S_{t}(r,\vartheta)-\Psi(r)S_{r}(r,\vartheta)\right)-aS_{\varphi}(r,\vartheta)\right]H_{-}^{(1)}(r,\vartheta)\nonumber\\
    &+\partial_{\vartheta}H_{-}^{(2)}(r,\vartheta)+\frac{1}{4}\left[2\cot{\vartheta}+4\left(k\csc{\vartheta}-a\omega\sin\vartheta\right)-iS_{\vartheta}(r,\vartheta)-\left(aS_{t}(r,\vartheta)\sin\vartheta+S_{\varphi}(r,\vartheta)\csc{\vartheta}\right)\right]H_{-}^{(2)}(r,\vartheta)=0\,,\label{DiracTorsCompNewVar3}\\
    &r\sqrt{\Psi(r)}\,\partial_{r}H_{-}^{(2)}(r,\vartheta)+\frac{i}{4r\sqrt{\Psi(r)}}\left[4\left(ak-\omega r^2\right)-r^{2}\left(S_{t}(r,\vartheta)+\Psi(r)S_{r}(r,\vartheta)\right)-aS_{\varphi}(r,\vartheta)\right]H_{-}^{(2)}(r,\vartheta)\nonumber\\
    &-\partial_{\vartheta}H_{-}^{(1)}(r,\vartheta)-\frac{1}{4}\left[2\cot{\vartheta}-4\left(k\csc{\vartheta}-a\omega\sin\vartheta\right)-iS_{\vartheta}(r,\vartheta)+\left(aS_{t}(r,\vartheta)\sin\vartheta+S_{\varphi}(r,\vartheta)\csc{\vartheta}\right)\right]H_{-}^{(1)}(r,\vartheta)=0\,.\label{DiracTorsCompNewVar4}
\end{align}

Thereby, by means of the separability ansatz
\begin{equation}
    H^{(1)}_{+}(r,\vartheta)=R_{1}(r)S_{1}(\vartheta) \,, \quad H^{(2)}_{+}(r,\vartheta)=R_{2}(r)S_{2}(\vartheta) \,, \quad H^{(1)}_{-}(r,\vartheta)=R_{3}(r)S_{1}(\vartheta) \,, \quad H^{(2)}_{-}(r,\vartheta)=R_{4}(r)S_{2}(\vartheta)\,,
\end{equation}
the Dirac equation is separated into radial and angular parts
\begin{align}
    &r\sqrt{\Psi(r)}R'_{1}(r)+\frac{i}{r\sqrt{\Psi(r)}}\left(ak-\omega r^2\right)R_{1}(r)=\lambda R_{2}(r)\,,\label{rad1MasslessSol}\\
    &r\sqrt{\Psi(r)}R'_{2}(r)-\frac{i}{r\sqrt{\Psi(r)}}\left\{ak-\omega r^2-3N_{1}\kappa_{\rm s}r+\frac{a}{\Psi(r)}\left[\left(4\Psi(r)-3\right)\cos\vartheta+6r\Psi(r)F(r,\vartheta)\right]\right\}R_{2}(r)=\lambda R_{1}(r)\,,\label{rad2MasslessSol}\\
    &r\sqrt{\Psi(r)}R'_{3}(r)-\frac{i}{r\sqrt{\Psi(r)}}\left\{ak-\omega r^2+3N_{1}\kappa_{\rm s}r-\frac{a}{\Psi(r)}\left[\left(4\Psi(r)-3\right)\cos\vartheta+6r\Psi(r)F(r,\vartheta)\right]\right\}R_{3}(r)=\lambda R_{4}(r)\,,\label{rad3MasslessSol}\\
    &r\sqrt{\Psi(r)}R'_{4}(r)+\frac{i}{r\sqrt{\Psi(r)}}\left(ak-\omega r^2\right)R_{4}(r)=\lambda R_{3}(r)\,,\label{rad4MasslessSol}
\end{align}
and
\begin{align}
    &S'_{2}(\vartheta)+\frac{1}{2}\left[\cot{\vartheta}+2\left(k\csc{\vartheta}-a\omega\sin\vartheta\right)\right]S_{2}(\vartheta)=-\,\lambda S_{1}(\vartheta)\,,\label{angular1MasslessSol}\\
    &S'_{1}(\vartheta)+\frac{1}{2}\left[\cot{\vartheta}-2\left(k\csc{\vartheta}-a\omega\sin\vartheta\right)\right]S_{1}(\vartheta)=\lambda S_{2}(\vartheta)\,,\label{angular2MasslessSol}
\end{align}
provided that the function $F(r,\vartheta)$ takes the form
\begin{equation}
    F(r,\vartheta) = \frac{\left(3-4\Psi(r)\right)}{6r\Psi(r)}\cos\vartheta+\frac{f(r)}{r}\,,\label{FixedFunction}
\end{equation}
where $f(r)$ is an arbitrary function of the radial coordinate and $\lambda$ is a separation parameter.

\section{Near-horizon solutions of the Dirac equation}\label{sec:sol}

Taking into account the specific form of the function~\eqref{FixedFunction} that ensures the separability of the Dirac equation and evaluating Eqs.~\eqref{rad1MasslessSol}-\eqref{rad4MasslessSol} in the near-horizon region, the system of radial equations expressed in tortoise coordinates reduces to leading order as
\begin{align}
    &R'_{1}(r^{*})+i\left(\frac{ak}{2mr_{h}}-\omega\right)R_{1}(r^{*})=0\,,\label{rad1MasslessHor}\\
    &R'_{2}(r^{*})-i\left(\frac{ak}{2mr_{h}}-\omega-\frac{3N_{1}\kappa_{\rm s}}{r_{h}}+\frac{3af(r^{*})}{mr_{h}}\right)R_{2}(r^{*})=0\,,\label{rad2MasslessHor}\\
    &R'_{3}(r^{*})-i\left(\frac{ak}{2mr_{h}}-\omega+\frac{3N_{1}\kappa_{\rm s}}{r_{h}}-\frac{3af(r^{*})}{mr_{h}}\right)R_{3}(r^{*})=0\,,\label{rad3MasslessHor}\\
    &R'_{4}(r^{*})+i\left(\frac{ak}{2mr_{h}}-\omega\right)R_{4}(r^{*})=0\,,\label{rad4MasslessHor}
\end{align}
whose general solution is
\begin{align}
    R_{1}(r^{*})&=B_{1}\,e^{i\Omega_{0} r^{*}}\,,\label{0thsol1}\\
    R_{2}(r^{*})&=A_{2}\,e^{-\,i\Omega_{+} r^{*}}\,,\\
    R_{3}(r^{*})&=A_{3}\,e^{-\,i\Omega_{-} r^{*}}\,,\label{0thsol3}\\
    R_{4}(r^{*})&=B_{4}\,e^{i\Omega_{0} r^{*}}\,,
\end{align}
where
\begin{align}
    &\Omega_{0} = \omega-\frac{ak}{2mr_{h}}\,, \quad \Omega_{\pm} = \omega-\frac{ak}{2mr_{h}}\pm\frac{3}{r_{h}}\left(N_{1}\kappa_{\rm s}-\frac{a}{m}f(r_{h})\right),
\end{align}
and $A_{2}$, $A_{3}$, $B_{1}$ and $B_{4}$ are constant parameters. Focusing then on the ingoing modes, the first contributions of these modes to the radial functions $R_{1}(r^{*})$ and $R_{4}(r^{*})$ in the near-horizon region are described by the radial equations
\begin{align}
    R'_{1}(r^{*})-i\Omega_{0}R_{1}(r^{*})&=A_{2}\frac{\lambda\sqrt{\Psi(r^{*})}}{r_{h}}\,e^{-\,i\Omega_{+} r^{*}}\,,\\
    R'_{4}(r^{*})-i\Omega_{0}R_{4}(r^{*})&=A_{3}\frac{\lambda\sqrt{\Psi(r^{*})}}{r_{h}}\,e^{-\,i\Omega_{-} r^{*}}\,,
\end{align}
which provides solutions as
\begin{align}
    R_{1}(r^{*})&=A_{2}\frac{i\lambda\sqrt{\Psi(r^{*})}\,e^{-\,i\Omega_{+}r^{*}}}{r_{h}\left(\Omega_{0}+\Omega_{+}\right)}\,,\\
    R_{4}(r^{*})&=A_{3}\frac{i\lambda\sqrt{\Psi(r^{*})}\,e^{-\,i\Omega_{-}r^{*}}}{r_{h}\left(\Omega_{0}+\Omega_{-}\right)}\,.
\end{align}

Thereby, the four radial solutions of the ingoing modes near the event horizon take the form:
\begin{align}
    R_{1}(r^{*})&=A_{2}\frac{i\lambda\sqrt{\Psi(r^{*})}\,e^{-\,i\Omega_{+}r^{*}}}{r_{h}\left(\Omega_{0}+\Omega_{+}\right)}\,,\\
    R_{2}(r^{*})&=A_{2}\,e^{-\,i\Omega_{+} r^{*}}\,,\\
    R_{3}(r^{*})&=A_{3}\,e^{-\,i\Omega_{-} r^{*}}\,,\\
    R_{4}(r^{*})&=A_{3}\frac{i\lambda\sqrt{\Psi(r^{*})}\,e^{-\,i\Omega_{-}r^{*}}}{r_{h}\left(\Omega_{0}+\Omega_{-}\right)}\,,
\end{align}
whereas the corresponding angular solutions to Eqs.~\eqref{angular1MasslessSol}-\eqref{angular2MasslessSol} are given by the polar-dependent part of the spin-weighted spheroidal harmonics~\cite{Newman:1966ub,Goldberg:1966uu,Teukolsky:1973ha,breuer1977some,Berti:2005gp}. Hence, in contrast to the standard case of GR, the helicity states of the Dirac fermion experience opposite energy shifts as a result of their interaction with the axial mode of torsion.

\section{Energy extraction without wave amplification}\label{sec:results}

Substituting the near-horizon solutions into the radial component of the Dirac current yields
\begin{equation}
    J^{r}(r,\vartheta)=\frac{1}{2r^{2}}\left[\left(|R_{1}(r)|^{2}-|R_{3}(r)|^{2}\right)|S_{1}(\vartheta)|^{2}-\left(|R_{2}(r)|^{2}-|R_{4}(r)|^{2}\right)|S_{2}(\vartheta)|^{2}\right],
\end{equation}
namely
\begin{equation}
    J^{r}_{h} (\vartheta) \approx -\,\frac{|A_{3}|^{2}|S_{1}(\vartheta)|^{2}+|A_{2}|^{2}|S_{2}(\vartheta)|^{2}}{4mr_{h}}\,,
\end{equation}
which provides a strictly positive net number current flowing down the black hole:
\begin{equation}
    \frac{dN}{dt} = -\,2mr_{h}\int_{0}^{2\pi}\int_{0}^{\pi}J^{r}_{h}(\vartheta)\sin\vartheta\,d\vartheta d\varphi \approx \frac{1}{4}\left(|A_{3}|^{2}+|A_{2}|^{2}\right),
\end{equation}
where the following normalisation condition has been used:
\begin{equation}
    \int_{0}^{\pi}|S_{1}(\vartheta)|^{2}\sin\vartheta \,d\vartheta = \int_{0}^{\pi}|S_{2}(\vartheta)|^{2}\sin\vartheta \,d\vartheta = \frac{1}{4\pi}\,.
\end{equation}

Thereby, as in the case of GR, it is clear that there is no net flux of Dirac fermions from the horizon, which agrees with the Pauli exclusion principle that limits the maximal occupation number of each mode.

On the other hand, the weak-energy condition is violated near the event horizon in the low-frequency regime. Specifically, considering for instance the timelike vector
\begin{equation}
    t^{\mu}=\frac{1}{r^{2}\sqrt{\Psi(r)}}\left(r^{2},0,0,a\right),
\end{equation}
the respective timelike projection of the energy-momentum tensor of the Dirac field reads
\begin{equation}
    T_{\mu\nu}t^{\mu}t^{\nu} = E_{1}(r,\vartheta)+E_{2}(r,\vartheta)+E_{3}(r,\vartheta)\,,
\end{equation}
where
\begin{align}
    E_{1}(r,\vartheta) =&\; \frac{\left(\omega r^{2}-ak\right)}{2r^4\Psi(r)}\left[\left(|R_{1}(r)|^{2}+|R_{3}(r)|^{2}\right)|S_{1}(\vartheta)|^{2}+\left(|R_{2}(r)|^{2}+|R_{4}(r)|^{2}\right)|S_{2}(\vartheta)|^{2}\right],\\
    E_{2}(r,\vartheta) =&\; \frac{a}{4r^{4}\sqrt{\Psi(r)}}\Bigl\{\sqrt{\Psi(r)}\left[\left(|R_{1}(r)|^{2}+|R_{3}(r)|^{2}\right)|S_{1}(\vartheta)|^{2}-\left(|R_{2}(r)|^{2}+|R_{4}(r)|^{2}\right)|S_{2}(\vartheta)|^{2}\right]\cos\vartheta\nonumber\\
    &-\left[\left(R_{1}(r)\bar{R}_{2}(r)+R_{3}(r)\bar{R}_{4}(r)\right)S_{1}(\vartheta)\bar{S}_{2}(\vartheta)+\left(\bar{R}_{1}(r)R_{2}(r)+\bar{R}_{3}(r)R_{4}(r)\right)\bar{S}_{1}(\vartheta)S_{2}(\vartheta)\right]\sin\vartheta\Bigr\}\,,\\
    E_{3}(r,\vartheta) =&\;\frac{3\left(N_{1}\kappa_{\rm s}r-2af(r)\right)}{4r^{4}\Psi(r)}\left[\left(|R_{1}(r)|^{2}-|R_{3}(r)|^{2}\right)|S_{1}(\vartheta)|^{2}+\left(|R_{2}(r)|^{2}-|R_{4}(r)|^{2}\right)|S_{2}(\vartheta)|^{2}\right].
\end{align}
In particular, in the near-horizon region this quantity takes the form
\begin{equation}
    T_{\mu\nu}t^{\mu}t^{\nu} \approx \frac{|A_{3}|^2|S_{1}(\vartheta)|^2\left(\Omega_{0}+\Omega_{-}\right)+|A_{2}|^2|S_{2}(\vartheta)|^2\left(\Omega_{0}+\Omega_{+}\right)}{4r^{2}_{h}\Psi(r)}\,.
\end{equation}
Hence, the weak-energy condition is strictly negative near the horizon if
\begin{equation}
    \omega < \frac{ak}{2mr_{h}}+\frac{1}{4}\,\gamma\left(\Omega_{+}-\,\Omega_{-}\right), \quad \gamma =\frac{|A_{3}|^{2}-|A_{2}|^{2}}{|A_{3}|^{2}+|A_{2}|^{2}}\,, \quad -\,1 \le \gamma \le 1\,,
\end{equation}
thus being violated by the Dirac fermion in a frequency regime that is shifted with respect to the case of GR by a factor $(\gamma/4)\left(\Omega_{+}-\,\Omega_{-}\right)$ in the presence of torsion.

In any case, it is worthwhile to stress that these results do not directly prevent the energy extraction from the rotating black hole. Indeed, a lengthy but straightforward calculation shows that unlike the classical case of GR the chiral asymmetry mediated by torsion gives rise to ingoing modes with negative energy that can reverse the sign of the energy current down the black hole
\begin{equation}
    \frac{dE}{dt} = -\int_{0}^{2\pi}\int_{0}^{\pi}T^{r}{}_{t}(r_{h},\vartheta)\sin\vartheta\,d\vartheta d\varphi\,,\label{Eflux}
\end{equation}
where
\begin{align}
    T^{r}{}_{t}(r_{h},\vartheta) = &-\frac{1}{16mr^{3}_{h}}\bigl\{r^{2}_{h}\bigl[\bigl(4\omega+\Omega_{-}-\,\Omega_{+}\bigr)|A_{3}|^2|S_{1}(\vartheta)|^2+\bigl(4\omega+\Omega_{+}-\,\Omega_{-}\bigr)|A_{2}|^2|S_{2}(\vartheta)|^2\bigr]\nonumber\\
    &+a\left(|A_{2}|^2|S_{2}(\vartheta)|^2-|A_{3}|^2|S_{1}(\vartheta)|^2\right)\cos\vartheta-2ak\left(|A_{3}|^2|S_{1}(\vartheta)|^2+|A_{2}|^2|S_{2}(\vartheta)|^2\right)\nonumber\\
    &-a\lambda\left(|A_{2}|^2+|A_{3}|^2\right)\left(S_{1}(\vartheta)\bar{S}_{2}(\vartheta)+S_{2}(\vartheta)\bar{S}_{1}(\vartheta)\right)\sin\vartheta\bigr\}\,.\label{Emom}
\end{align}

By taking into account the following properties of the spin-weighted spheroidal harmonics:
\begin{align}
    K_2 - 2k I_2 + 2a\omega L_2 &= \lambda J\,,\\
    K_1 + 2k I_1 - 2a\omega L_1 &= -\,\lambda J\,,
\end{align}
with
\begin{align}
    I_i &:= \int_{0}^{\pi} |S_i(\vartheta)|^2 \sin\vartheta\, d\vartheta\,, \quad K_i := \int_{0}^{\pi} |S_i(\vartheta)|^2 \cos\vartheta\, \sin\vartheta\, d\vartheta\,,\quad L_i := \int_{0}^{\pi} |S_1(\vartheta)|^2 \sin^3\vartheta\, d\vartheta\,, \quad i=1,2\,,\\
    J &:= \int_{0}^{\pi}
    \bigl(S_1(\vartheta)\bar{S}_2(\vartheta)+\bar{S}_1(\vartheta)S_2(\vartheta)\bigr)
    \sin^2\vartheta\, d\vartheta\,,
\end{align}
expression~\eqref{Eflux} takes the following form in the slow rotation approximation:
\begin{equation}
    \frac{dE}{dt} = \frac{1}{32mr_{h}}\left[\bigl(4\omega+\Omega_{-}-\,\Omega_{+}\bigr)|A_{3}|^2+\bigl(4\omega+\Omega_{+}-\,\Omega_{-}\bigr)|A_{2}|^2\right].
\end{equation}

Thus, the energy current becomes negative for modes with frequencies in the range:
\begin{equation}
    \omega < \frac{1}{4}\,\gamma\left(\Omega_{+}-\Omega_{-}\right), \quad -\,1 \le \gamma \le 1\,,
\end{equation}
indicating an outward flow of energy from the horizon.

\section{Conclusions}\label{sec:conclusions}

In this work, we have analysed the behaviour of Dirac fermions in the vicinity of a rotating black hole with torsion. Specifically, we have considered a black hole solution recently obtained in the framework of cubic Poincaré gauge theory, which includes a spin charge as a source of torsion and displays a spin-orbit interaction in the gravitational action~\cite{Bahamonde:2025fei}. Indeed, these two quantities are encoded in the axial mode of torsion, which makes this analysis particularly relevant, since the interaction of Dirac fermions minimally coupled to torsion is mediated exclusively by this mode~\cite{Hehl:1971qi}.

Computing the explicit form of the Dirac equation for a massless fermion in this space-time, we have determined the condition that the general function describing the spin-orbit interaction of the solution must satisfy to guarantee separability. Thereby, this condition has allowed us to obtain analytical solutions to this equation in the near-horizon region, which present helicity states with opposite energy shifts induced by the axial mode of torsion.

From these results, it is then straightforward to derive the corresponding net number and energy currents associated with the ingoing modes of the Dirac field. In particular, the net number current acquires strictly positive values near the horizon, indicating the absence of wave amplification, in agreement with the Pauli exclusion principle. Furthermore, as in the standard case of GR, the weak-energy condition can also be violated in the near-horizon region by the Dirac fermion in a frequency range that is determined by the angular velocity of the solution, as well as by the spin charge and the spin-orbit interaction related to torsion. However, the energy current takes negative values for a specific range of frequencies, thus signaling the energy extraction by the Dirac fermion from the rotating black hole.

Therefore, our analysis reveals that Dirac fermions can extract energy from the black hole even though the ingoing modes do not undergo amplification. In this regard, it is important to stress that no superradiant regime is observed in the usual sense, but the energy flux indicates a net transfer of energy away from the black hole. This demonstrates that energy extraction can occur independently of wave amplification and should not be regarded as synonymous with superradiance, highlighting that the range of mechanisms through which black holes can lose energy is broader than traditionally assumed.

\noindent
\section*{Acknowledgements}

We would like to thank José M. M. Senovilla for helpful discussions. The work of S.B. is supported by the Institute for Basic Science (IBS-R018-D3). The work of J.G.V. is supported by the Institute for Basic Science (IBS-R003-D1).

\newpage

\bibliographystyle{utphys}
\bibliography{references}

\end{document}